# Systematic Study of Nuclear Gamma-Ray Spectra of One Hundred Super Novae Expected by Future Nuclear Gamma-Ray Imaging Spectroscopic Observations

Yoshitaka Mizumura[1,2], Atsushi Takada[1], and Toru Tanimori[1]

1 Graduate School of Science, Kyoto University, Sakyo, Kyoto 606-8502, Japan
2 Unit of Synergetic Studies for Space, Kyoto University, Sakyo, Kyoto, 606-8502, Japan

Abstract

Supernovae (SNe) are the most fascinating objects in astronomy and are intensely investigated. However, many mysteries such as nucleosynthesis and the origin of SNe Ia remain unsolved. Although the thermonuclear explosion of a single-degenerate white dwarf has been considered to be the origin of SNe Ia, a merger of two white dwarfs (double-degenerate scenario) has been frequently denoted to be more promising than a single-degenerate white dwarf. Recently the importance of observing the MeV gamma-ray band to conclusively determine the origin has been remarked. MeV gamma-rays are unique probes directly emitted from the exploding or merging region. It is evident that statistical analysis based on imaging spectroscopic observations of ~100 SNe Ia with MeV gamma-rays is necessary to obtain a definite answer. To achieve this, a telescope with a sensitivity that is 100 times that of COMPTEL is necessary. Proper imaging spectroscopy for the MeV gamma-ray band has been established by an electron-tracking Compton camera; hence, a concrete design of a MeV gamma-ray telescope has been proposed in our previous work. We have studied the details of the spectroscopic feature of SNe Ia based on the performance of a proposed telescope and found that statistical analysis can considerably suppress fluctuations of the individual properties of SNe and reveal their intrinsic differences in averaged light curves of SNe up to 60 Mpc. Our answer for the origin of SNe Ia extends to the case of single-degenerate scenario and double-degenerate coexistence scenario.

Introduction

Supernovae (SNe) Ia are known to be one of the most interesting objects in astronomy owing to their important role such as standard candles for measuring cosmological distances (Tammann & Leibundgut 1990; Riess et al. 1998), factories of nucleosynthesis (Bertulani & Kajino 2016), sources of kinetic energy in the galaxy evolution process (Powell et al. 2011), accelerators of galactic cosmic rays (Helder et al. 2009), galactic positron sources (Prantzos et al. 2011), and terminuses of stellar binary evolution (Postnov & Yungelson 2014). There is a general consensus regarding the properties of the primary star that is associated with thermonuclear explosions of carbon–oxygen white dwarfs (WDs) near the Chandrasekhar mass in close binaries (Bloom et al. 2012). However, the nature of the binary companion and the manner in which it leads the WDs to mass growth, ignition, and explosion is still poorly understood (Howell 2011; Maoz et al. 2014). In contrast to the case of core-collapse SNe (Smartt 2009), there are few observations that help to identify the SN Ia progenitors from surveys in either the companions remaining after explosions or progenitor stars in pre-explosion images (Maoz & Mannucci 2008).

The following are the two leading scenarios that are widely discussed regarding the nature of the progenitor systems (Maeda & Terada 2016): the single-degenerate (SD) scenario (Whelan & Iben 1973) and the double-degenerate (DD) scenario (Iben & Tutukov 1984; Webbink 1984). In the SD scenario, a WD accretes mass from a close companion through Roche lobe overflow or stellar wind until it reaches the Chandrasekhar limit. In the DD scenario, a close binary system of two WDs loses angular momentum through the radiation of gravitational waves, until the two WDs finally merge. Major research efforts—not only observational but also experimental and theoretical—have been made to try and solve the "SN Ia progenitor problem". However, this problem is still open to discussion. Identification of progenitor systems and study of the population of progenitor systems have ramifications for cosmology, the evolution of galaxies, SN explosion models, and binary evolution theories (Wang & Han 2012; Maoz & Mannucci 2012).

The optical light curves of SNe Ia are powered by the radioactive decay chain of $^{56}$Ni. Typically, it is expected that a SN Ia explosion produces 0.5–0.6 $M_{solar}$ of $^{56}$Ni (Nomoto et al. 1984). It decays into $^{56}$Co and subsequently into stable $^{56}$Fe, which have half-lifetimes of 6.1 days and 77.2 days, respectively (Nadyozhin 1994; Junde et al. 2011). Although SN Ia is accepted as a nuclear gamma-ray emitter associated with the $^{56}$Ni chain, only upper limits on the gamma-ray emission have been observed since the 1980s (Horiuchi & Beacom 2010) until recently.

The MeV gamma-ray light curve is expected to be a promising tool to distinguish between the progenitor scenarios (i.e., SD versus DD) (Horiuchi & Beacom 2010; Summa et al. 2013). In 2014, direct measurement of MeV gamma-rays was performed successfully via SPI/INTEGRAL for the first time on a SN Ia, SN2014J (Churazov et al. 2014; Diehl et al. 2014), which is the closest SN Ia at a distance of 3.53 Mpc (Karachentsev & Kashibadze 2006) observed in the past four decades (Zheng et al. 2014), and possibly the closest in the past 130 years (Crotts 2015). Although SPI has a large effective area (several tens of $cm^2$) (Attié et al. 2003), the observed light curves are of insufficient quality to determine the progenitor scenario (Churazov et al. 2015). In addition, HXD/Suzaku also attempted to measure soft gamma-rays from SN2014J; however, only a signal with a significance level of 2 σ has been reported (Terada et al. 2016). Both SPI and HXD are incapable of proper imaging spectroscopy (Vedrenne et al. 2003; Takahashi et al. 2007) and have consequently suffered from intense background radiation, which is one of the most difficult problems in MeV gamma-ray astronomy (Weidenspointner et al. 2001, 2005; Schönfelder 2004). A proper imaging spectroscopy that focuses a half of the gamma-rays within 2° radii, with the same detection area as that of SPI, would easily improve the data quality. In the case of the 847 keV line observation of SN2014J by SPI, the detection significance of ~4 σ will be improved to ~100 σ. However, a proper imaging spectroscopic method for Compton scattering gamma-rays such as this was not realized until the advent of the electron-tracking Compton camera (ETCC) (Tanimori et al. 2004; Tanimori et al.2015).

A standard Compton camera (CC) is frequently used as a gamma-ray imager in various fields. However, CC is incapable of conducting proper imaging as it only measures a polar angle of the two incident angles of gamma-rays. It can only restrict the incident gamma-ray direction as an annulus on the field of view (FoV). Although some imaging optimization algorithms have been expected to complement the lack of one incident angle of gamma-rays in CC, there is absolutely no such a method to improve the statistics as indicated in our previous works (Tanimori et al. 2015; Tanimori et al. 2017).

Recent proposals of MeV missions based on CCs advocate the employment of an angular resolution measure (ARM) as a half-power radius (HPR) or point-spread function (PSF). However, the ARM is only the resolution of the Compton scattering angle and does not represent the angular resolution of a gamma-ray telescope. Therefore, such an approach is insufficient to estimate the sensitivity. Standard CCs provides an ARM of several degrees and an obscure HPR of several ten degrees (Tanimori et al. 2017). Their true sensitivities are degrade by more than an order of magnitude than that using an ARM as a HPR or PSF. It also produces a significant uncertainty, depending on the amount of background noise. An improvement of the ARM resolution on its own is not an efficient approach to improve the PSF of a gamma-ray telescope.

An ETCC has the proper imaging capability to process the incident gamma-ray photon by photon. This allows the measurement of complete directional information, i.e., both the polar and azimuth angles of Compton scattering. It is the first MeV gamma-ray telescope based on complete geometrical optics, which provides a well-defined PSF with a few degrees of the HPR and efficient background rejection (Tanimori et al. 2015; Tanimori et al. 2017; Mizumoto et al. 2015).

Once a well-defined PSF and the effective area and flux of the background are determined for the instrument, the sensitivity can be uniquely and simply determined. We revealed a concrete model of a satellite mission equipped with four modules of $(50\ \text{cm})^3$-cubic ETCCs (ETCC satellite). It would reach a sensitivity beyond 1mCrab with a HPR of 2 degrees and an effective area of ~200 cm$^2$ (Tanimori et al. 2015). The effective area and HPR seems realizable in the near future since this model was estimated only using the present technology of radiation detectors. In addition to the well-defined PSF, ETCC provides several tools (a light material scatterer, particle identification using dE/dx of tracks, and kinematical testing) for the efficient rejection of all types of backgrounds such as charged cosmic rays, neutrons, and chance coincidences, which include simultaneous multi-hitting gamma-rays. The ETCC's sophisticated examination of events would certainly reduce backgrounds down to the cosmic diffuse gamma-ray flux, which is intrinsic and can never be removed.

There is a possibility that we might reach or exceed the 1mCrab flux level sensitivity in the MeV gamma-ray band in the near future. Therefore, we must consider possible new approaches to astrophysics. In this paper, we show the behaviors of the ETCC's angular related parameters and the expected results of imaging spectroscopic observations of SNe Ia in the MeV gamma-ray band with a 1mCrab flux level sensitivity. Such sensitive observations will enable us to begin a statistical analysis of spectroscopic features of SNe since nearly a hundred SNe would be observed during one satellite mission.

Instrument and Point-Spread Function

To estimate gamma-ray spectra from SNe Ia of both the SD and DD scenarios, which would be detected by ETCC satellite, we used the expected performances that are as follows: an effective area of 240 cm$^2$ at 0.8 MeV, an energy resolution of $\Delta E/E = 5.0 \times (E/662\ \text{keV})^{-0.5}$ [%] (FWHM), a high live-time ratio of ~1/3 (owing to a large FoV of $2\pi$ sr), and the angular resolution shown in Fig. 2. All of these figures are derived by simply extending the present performances of SMILE-II (Mizumura et al. 2014; Matsuoka et al. 2015; Tanimori et al. 2015; Takada et al. 2016) and existing technologies. One of key improvements made on the ETCC, in comparison to the SMILE-II, is an alignment of the pixel scintillator arrays (PSA) inside the vessel of a gaseous time projection chamber (TPC). Previously, the PSAs were set outside the gas vessel of the TPC. This change allows the PSAs to detect high energy (above several hundred keV) Compton-recoil electrons that have escaped from the TPC. Additionally, the problem of the deterioration of angular resolution of a recoil electron is resolved by measuring both points of an electron track in the TPC and in the PSAs. Thus, it improves both the effective area and angular resolution of a recoil electron significantly. This improvement was certificated by the SMILE-II+ balloon experiment (Takada et al. 2016) performed in April 2018.

The possibility of a sharp PSF is key to reaching a 1mCrab flux sensitivity with high reliability. In accordance with the usage in MeV gamma-ray astronomy, in this study, we use the conventional angular resolutions of CC and ETCC: ARM for the direction of the scattered gamma-ray and scatter plane deviation (SPD) for the direction of a recoil electron along the scattering cone in the Compton scattering process. Although ARM and SPD are useful for evaluating detector performance, PSF is not obtained simply from these resolutions. ETCC can provide a comparable quality of ARM with that of advanced CC (~5° at 662 keV) and can also provide a unique SPD by measuring the track of the Compton-recoil electron. Although the SPD of the current ETCC is still moderate (several ten degrees), we are developing it to realize an SPD of several degrees limited by the multiple scattering of a recoil electron in the gas. For low-energy electrons, the installation of third electrodes in a micro-pattern gaseous detector may possibly improves the SPD, and eventually pixel readout technology would improves it in the near future. Then, we estimate the SPD determined at a distance of 2.4 mm (three data points with 0.8 mm pitch readout) from the scattering, considering only the multiple electron scattering in the gas.

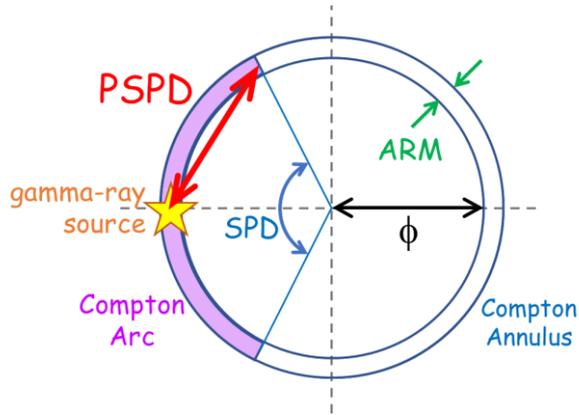

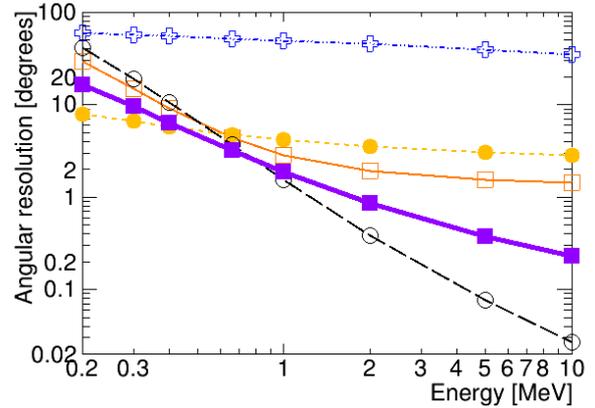

**Fig. 1**, Schematic explanation for projected SPD (PSPD) in the sky. A conventional Compton camera reconstructs gamma-rays as a Compton circle with the radius of the Compton scattering angle ϕ. Compton annulus is the error region of the Compton circle, which has a width consistent with that of ARM. Compton arc has a more reduced error region compared to that of the Compton annulus. This is achieved using the recoil-direction of the Compton electron. SPD is a resolution of the recoil-direction of an electron defined on the Compton circle. PSPD is the maximum angular distance of the Compton arc region from the gamma-ray source which is defined in the sky. Thus, we can evaluate the angular resolutions of two angles of gamma-ray using ARM and PSPD, both of which are defined in the sky.

SPD does not correspond to the angular dispersion of incident gamma-rays in the sky since it is defined on the Compton circle, as shown in Fig. 1. We introduce the projected SPD (PSPD) for comparison to angular resolution related parameters. PSPD is defined as the maximum angular distance from the gamma-ray origin in the sky, and it is smaller than the diameter (2ϕ) of the Compton circle anytime, as described in Fig. 1. We show the variations of ARM, PSPD, and PSF as a function of the energy of the incident gamma-rays in Fig. 2. The PSPD give a narrow window of error of ~0.1° less than that of PSF for gamma-rays with energies of several MeV. If we know the precise position of the source, such as a position-identified SNe Ia, we can apply stricter imaging selection than simple radius cut with HPR. Figure 3 shows a schematic of imaging selection windows, which are rectangular windows made of a combination of ARM and SPD. Contamination of the background can be assumed to be proportional to the solid angles of imaging selection windows, assuming isotropic background. Therefore, the background suppression of the rectangular window cut is expressed by solid angles or an equivalent angular radius of a circular region that has the same solid angle as those of the rectangular window. The factor of background suppression by the rectangular window in comparison to the HPR is roughly 10, and this improves the sensitivity of the faint signal by a factor of ~3.

**Fig. 2**, Energy dependences of angular resolution related parameters. The filled circles and open circles indicate ARM and PSPD, respectively. PSF and equivalent angular radius of rectangular window are shown as open squares and filled squares, respectively. The open crosses are PSF without SPD information (i.e., conventional Compton mode).

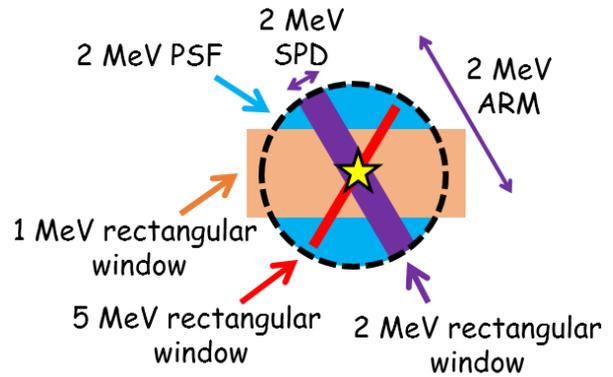

**Fig. 3**, Schematic for explaining PSF (2 MeV) and rectangular windows (1, 2, and 5 MeV). The star indicates gamma-ray source position. The orange, purple, and red rectangle regions are rectangular windows for 1, 2, and 5 MeV photon, respectively. Typically, SPD is narrower than ARM for high energy photons. PSF region for a 2 MeV photon is shown as the dashed circle.

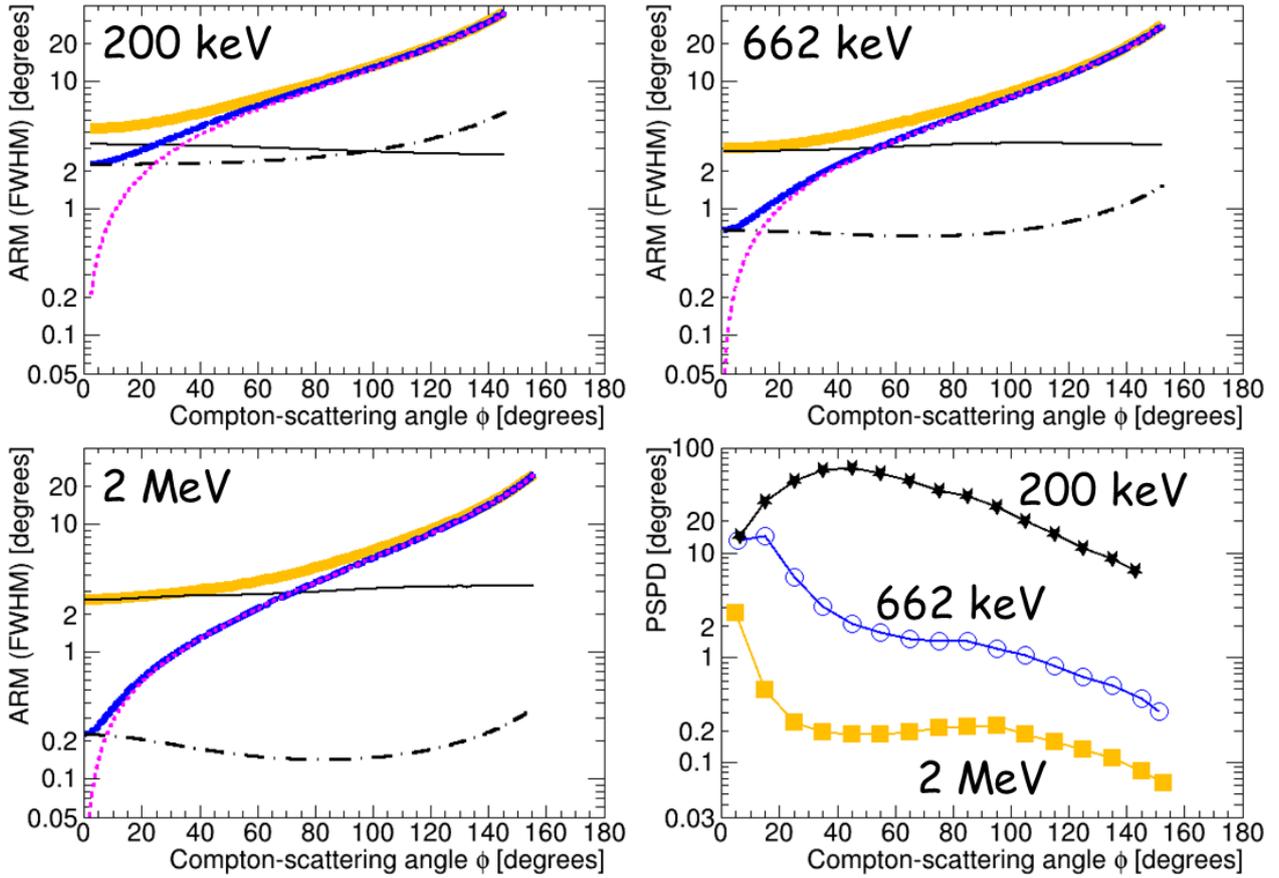

**Fig. 4**, Top-left, top-right, and bottom-left panel show ARM as a function of the Compton scattering angle ϕ with incident gamma-ray energies of 200, 662, and 2000 keV, respectively. In these panels, orange-thick lines, thin-solid lines, and purple-bold lines indicate total ARM resolutions, contribution of position resolution, and contribution of energy resolution, respectively. Furthermore, the component of energy resolution can be resolved to contributions of energy resolutions of Compton-recoil electrons and Compton-scattered gamma-rays, which are also shown as the dotted line and the dot-dashed lines, respectively. The bottom-right panel shows the Compton scattering angle dependence of PSPD with incident gamma-ray energies of 200, 662, and 2000 keV, which are plotted as black stars, blue open circles and orange squares, respectively.

For certain energies of the incident gamma-ray, Fig. 4 shows that ARM and PSPD vary considerably depending on the Compton scattering angle ϕ. Consequently, PSF is also a function of ϕ, as shown in Fig. 5. To maintain an ideal PSF for gamma-rays with low energies of less than 300 keV, events with only an electron with an energy higher than 50 keV can be used by considering the relatively intense flux of low-energy gamma-rays.

In addition, the large dependence of ARM on ϕ is shown in Fig. 4, in which ARM evidently decreases as ϕ increases. As ϕ decreases, contribution of position resolution—including resolution of the depth of the interaction (DOI)—becomes dominant for ARM. Therefore, the geometrical size of the absorber for a Compton-scattered gamma-ray is more essential than its energy resolution. Thus, the energy resolution of the absorber is less significant for obtaining a good ARM. Surprisingly no studies have been conducted to discuss this as the energy resolution of the absorber is considered to be the most significant factor in determining the performance of standard CCs. Furthermore, as the φ increases, ARM becomes intrinsically small, with a rough proportion to the ratio of the energy of the recoil electron to that of the scattered gamma-ray. For φ larger than ~60° for all energies in the sub-MeV and MeV ranges, ARMs are found to be intrinsically larger than ~5° even for expensive scintillators with energy resolutions of ~3%–5% (FWHM) at 662 keV (e.g., GAGG: Iwanowska et al. 2013, and LaBr$_3$: Quarati et al. 2011).

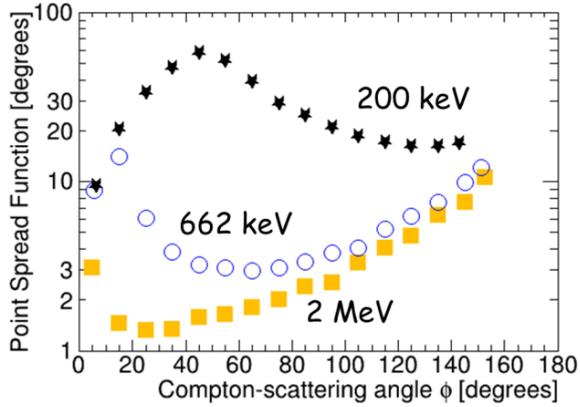

**Fig. 5**, PSF as function of Compton scattering angle φ with incident gamma-ray energies of 200, 662 and 2000 keV, which corresponded to the black filled stars, the blue open circles, and the orange filled squares, respectively.

Contrary to relation of energy resolution and ARM, the success of precise 3D tracking a Compton-recoil electron enable us to develop a thorough understanding of the imaging method based on the Compton process and consequently reveal quantitatively how to attain a good PSF in the Compton process. It is demonstrated that the improvement of SPD is more essential than that of ARM to attain a sensitivity beyond 1mCrab flux level in MeV gamma-ray astronomy.

Background Estimation

The estimation of the background intensity in the MeV region is also important for the estimation of a sensitivity. The intensity of celestial diffuse gamma-ray background and the contamination of non-gamma-ray origins have already been discussed in our previous work (Mizumura et al. 2017). The diffuse background spectra are compiled by the BAT/Swift team (Ajello et al. 2008), and we conservatively assumed an intensity two times higher than this diffuse background spectra. Now what we should consider to be backgrounds are gamma-rays generated in the instrument and non-gamma-ray origins such as cosmic rays, neutrons, and chance coincidence events.

To suppress these backgrounds efficiently, we have introduced a particle identification by the dE/dx of a Compton-recoil electron, kinematical testing using the angle between a Compton-scattered gamma-ray and a Compton-recoil electron, and the use of a low atomic number material as a scatterer. Results of the SMILE-I experiment, in which dE/dx was mainly used for background rejection, is a useful reference in determining the estimation of the backgrounds (Takada et al. 2011). After the analysis, ~900 gamma-ray events remained. These events contained diffuse cosmic gamma-rays, diffuse atmospheric gamma-rays and contributions resulting from non-gamma-ray origins, i.e., instrumental gamma-ray emissions. Using the variation of the flux of these components by the atmospheric depth, it was found that one half of the events could be attributed to cosmic diffuse gamma-rays, while the other half resulted from backgrounds. Therefore, the adoption of two-time high intensity diffuse gamma-ray backgrounds as the background is quite conservative since the fine dE/dx and kinematical cut in the recent development of SMILE would efficiently suppress background such as up-going gamma-rays and accidental events remaining in the final events of SMILE-I.

Simulation Results of Imaging Spectroscopy for SN Ia

A gamma-ray flux from SNe Ia was estimated as follows. Our simulation was conducted based on the theoretical results of Summa et al. (2013). Their work provided the gamma-ray spectrum of SNe Ia at a distance of 1 Mpc, which are used in this work as templates (hereafter referred to as SN-templates). Their work also indicated that the time evolution of angle-averaged spectra during the 100 days after the explosions is important to the diagnose SNe Ia progenitor scenarios. The following two scenarios are investigated: the SD scenario, in which a near Chandrasekhar mass white dwarf explodes as a delayed detonation, and the DD scenario, in which a 1.1 $M_{solar}$ and a 0.9 $M_{solar}$ WD violently merge. A detailed description of the calculation frameworks of the SN-templates are described in Summa et al. (2013) with references. The SN-templates provided snapshots of gamma-ray spectra at the follows times of 20.1, 34.9, 54.3, 75.7, and 101.7 days after the explosion in both the SD and DD scenario. Realistic gamma-ray spectra of SNe and instrumental performance are estimated using simulations through the following three steps. (1) Smearing of the SN-templates by the instrument's energy resolution. (2) Estimating the number of gamma-rays detected by the effective area of the instrument using the smeared SN-templates, and emulating the photon fluctuation according to Poisson statistics. (3) Extracting the number of gamma-rays from those detected from SNe within the PSF and estimating the background level within the region of interest. For this calculation, we used the rectangular window presented in Fig. 2.

We must mention the suppression of background fluctuation as it is one of important issues in the observation of SNe in the gamma-ray band (Terada et al. 2016). The wide FoV of ETCC is a great advantage for suppressing the background fluctuation, which enables us to measure a wide region of the background sky simultaneously. Therefore, we can efficiently suppress the observational fluctuation of the background.

Figure 6 shows the expected results of observable gamma-ray spectra in both SD and DD scenarios at a distance of 5, 20, 40, and 60 Mpc with accumulating gamma-rays within 30–160 days after the explosion. It can be observed that several line gamma-ray features clearly even a SN Ia exploding at a distance of 60 Mpc. For spectra beyond 40 Mpc, the photon fluctuation limit is dominant for the continuum flux level. Therefore, we try to suppress the fluctuation by averaging over several observations of SNe Ia. Within five years of operating the ETCC satellite, 27 and 90 SNe Ia can be expected to be observed at distances up to 40 and 60 Mpc, respectively, using the explosion rate of SNe Ia in the nearby universe with $\sim 2 \times 10^{-5}$ yr$^{-1}$ Mpc$^{-3}$ (Horiuchi & Beacom 2010; Maoz & Mannucci 2012, e.g.). We averaged the spectra of 27 and 90 SNe Ia at the distances of 40 and 60 Mpc as shown in Fig.7

To distinguish between the progenitor scenarios of SN Ia, we used gamma-ray light curves to observe SN Ia that exploded at a distance of 1 Mpc with energy ranges of 0.4–4.0, 0.7–4.0, and 1.0–4.0 MeV, as shown in the Fig. 8. The expectations are based on the interpolation of SN-templates 100 days after the explosion and it decays with life time of $^{56}$Co beyond 100 days since explosion, because gamma-ray light curves several ten days after the explosions are thought to be powered by $^{56}$Co. We adopt an energy range of 0.7–4.0 MeV for our studies of gamma-ray light curves because the energy range provides the best signal to noise ratio for SNe Ia at a distance beyond 20 Mpc. We must also consider the dependence of gamma-ray flux on the produced-mass of $^{56}$Ni (both SD and DD scenarios) and viewing angles of the progenitor binary system (only DD scenario). We adopt fluctuations of 20% for the produced-mass of $^{56}$Ni and 30% for the viewing angles of each SN Ia explosion (Maeda, K., private communication).

To study light curve diagnostics, we generate 90 SNe (expected number of SN Ia explosions up to 60 Mpc over five years) using random numbers. Each SN has individual properties such as explosion distance, relative intensity due to the production of $^{56}$Ni mass, and relative intensity due to the viewing angle of a WD binary. The properties of a generated SNe and its relations are shown in Fig. 9. Figure 10 shows the simulated light curves averaged over the five years of the ETCC satellite operation for SNe Ia with distance ranges of 0–20, 20–40, and 40–60 Mpc, which are scaled to the mean distance of each distance range. Light curves without any fluctuations are drawn by solid lines and all fluctuations that are considered are drawn by circles (SD scenario) and squares (DD scenario).

Discussion

Figure 6 indicates the simulated spectra of 5, 20, 40, and 60 Mpc SNe Ia for an accumulation time of 130 days, during which 18 SNe Ia are expected with a distance up to 60 Mpc in one year, including the observational duty factor. The bolometric detection integrating from 0.7–4.0 MeV is also possible for the distant SNe at 60 Mpc with >3 σ for an observation period of 10 days. We could obtain the light curve for MeV gamma-rays for ~20 SNe every year.

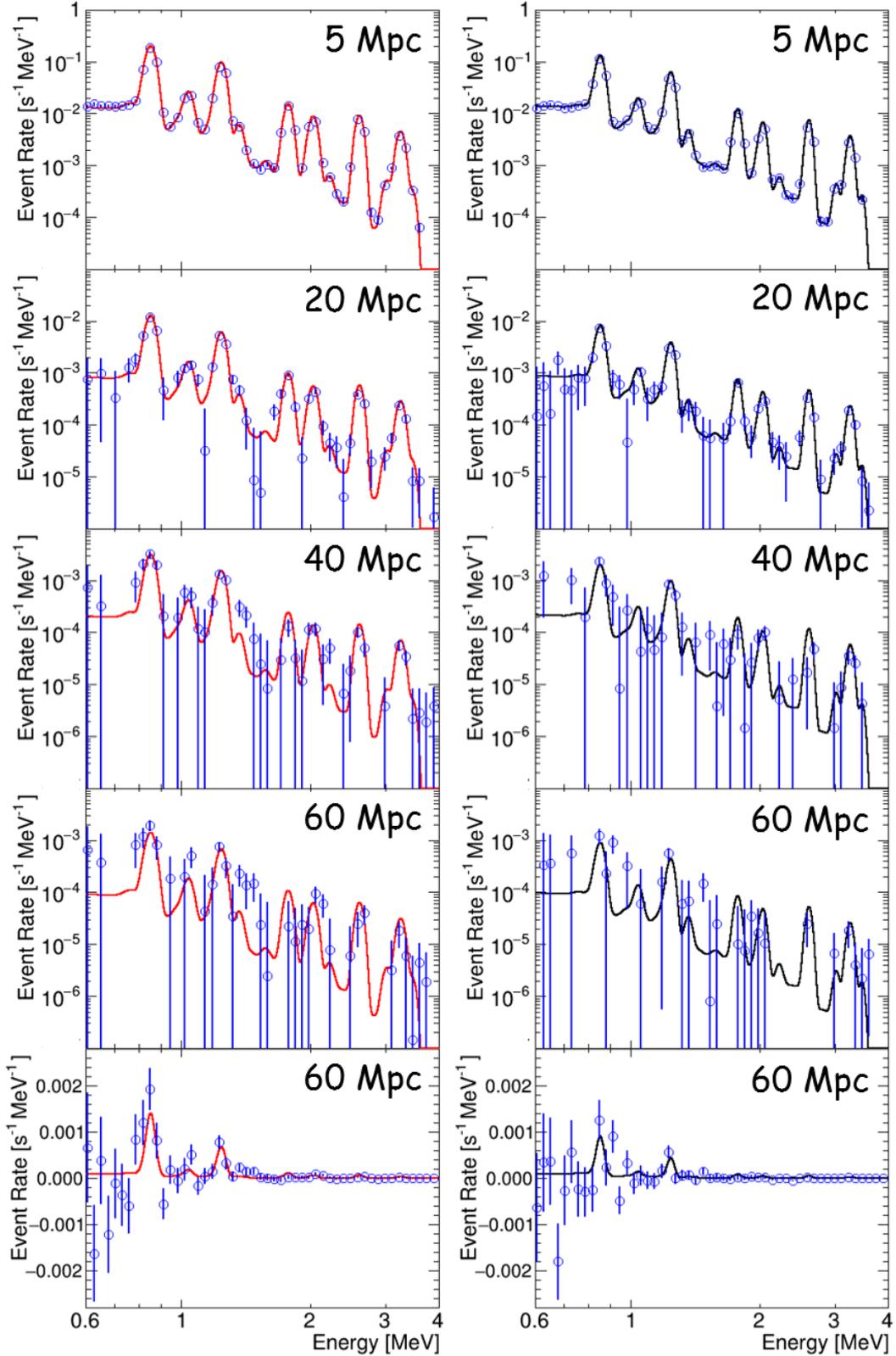

**Fig. 6**, Simulated SNe Ia spectra of both SD (left panels) and DD (right panels) scenarios exploding at distances of 5, 20, 40, and 60 Mpc. These are estimated using the rectangular window cut with observation between 30 to 160 days since the explosion. Observational duty ratio of 1/3 is considered, and the large duty ratio is realized by the wide FoV of ETCC. The open circles and solid lines indicate expected-observable spectra including photon fluctuations and model SN Ia spectra. The background has been subtracted from both of these spectra.

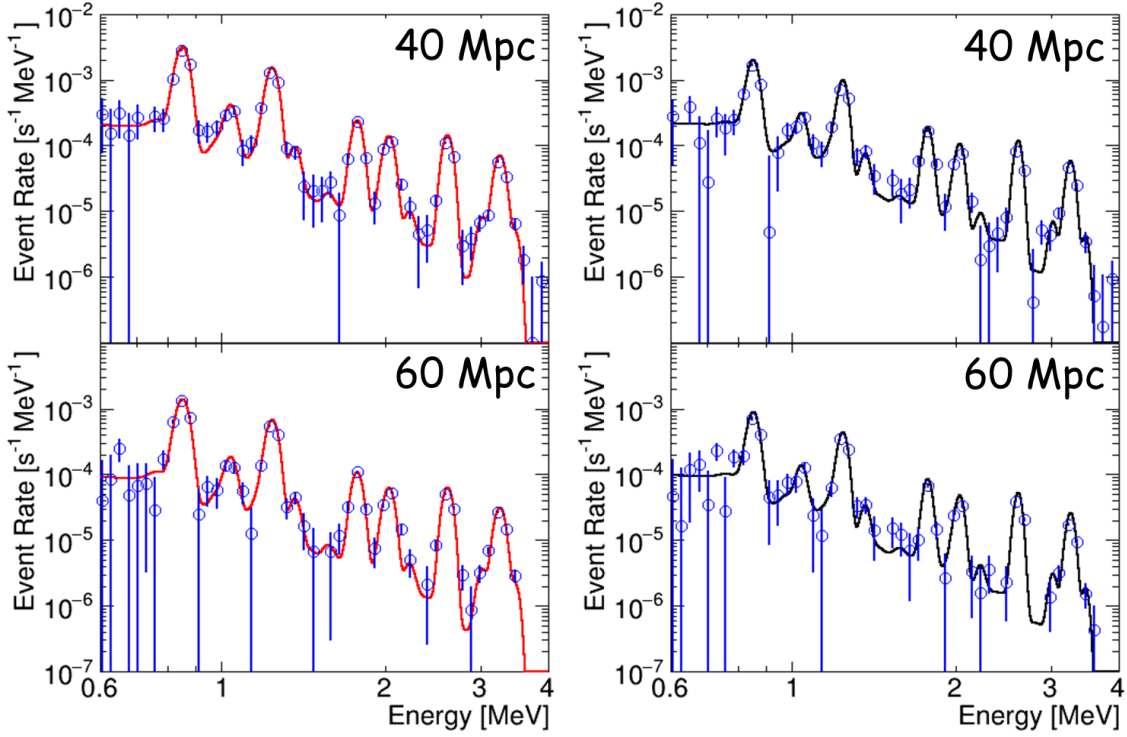

**Fig. 7**, Expected spectra of SNe Ia averaged over 27 and 90 SNe, which are expected number of SNe over five years of operation within 40 and 60 Mpc. The left panels and right panels indicate the SD and DD scenarios, respectively. The top panels and bottom panels are spectra of SNe Ia exploding at distances of 40 and 60 Mpc.

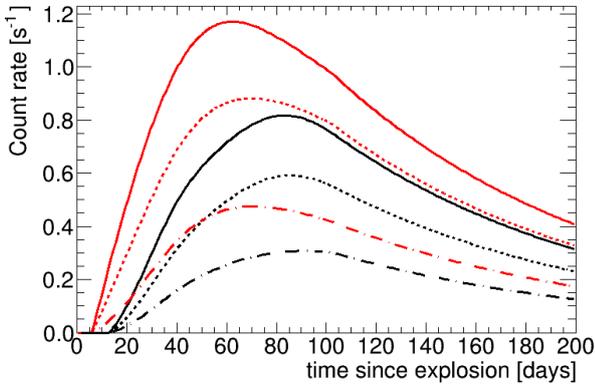

**Fig. 8**, Light curves of SN exploding at a distance of 1 Mpc. These curves are obtained by integrating over the count rate in the energy ranges of 0.4–4.0, 0.7–4.0, and 1.0–4.0 MeV. The red lines indicate the case of SD scenario and the black lines indicate the case of DD scenario. The solid lines, dotted lines, and dot-dashed lines show energy ranges of 0.4–4.0, 0.7–4.0, and 1.0–4.0 MeV, respectively. The light curves before 100 days after the explosion are smoothly interpolated using a spline curve, and the light curves after 100 days since explosion are considered to be decaying with the time scale of $^{56}$Co lifetime.

It is noted that the averaged light curves of both of the SD and DD scenarios in a distance range of 0–20 Mpc are evidently higher than the light curves without any fluctuations. This behavior is caused by one SN Ia that has a large intensity owing to the fluctuation of $^{56}$Ni mass production. Averaged light curves, except the high-intensity SN Ia, are plotted as blue open circles in Fig. 9. The data points are well-fitted to a light curve without fluctuations. On the other hand, averaged light curves without any selection for SNe Ia in both the SD and DD scenarios in distance ranges of 20–40 and 40–60 Mpc are fitted to curves without any fluctuations. This means that cancelling out the fluctuations in the properties of each SN Ia by averaging a few tens of SNe is quite important in the diagnostics of SN Ia progenitors. Consequently, the averaged light curves enable us to distinguish between SD and DD scenarios as the SN Ia progenitor.

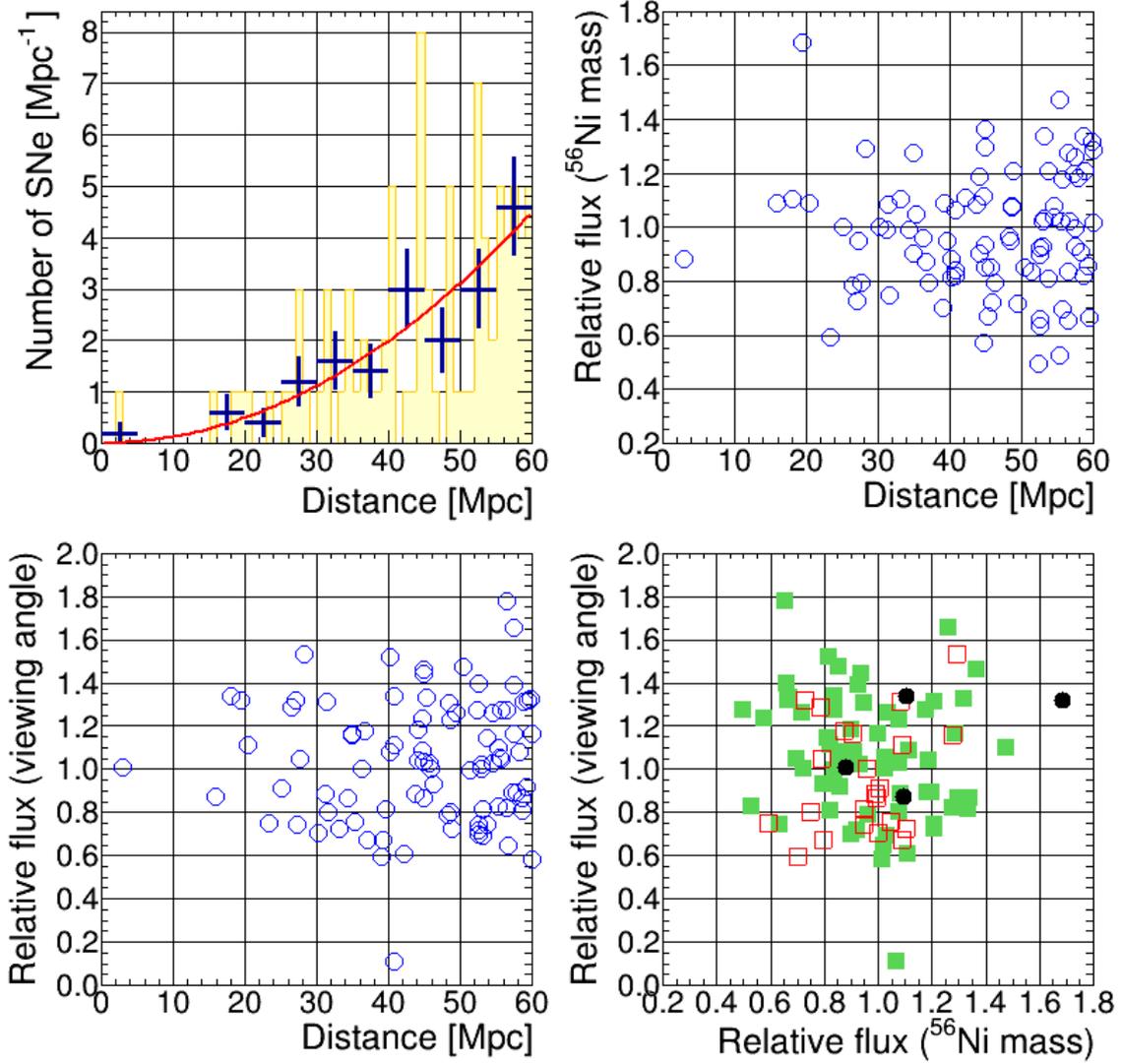

**Fig. 9**, Distributions of properties of individual SNe generated by random numbers. Exploding distances of the SNe are drawn in the left-top panel. The solid line is the assumed number density of a SN explosion with the function of distance. The histogram is a generated distance distribution by random numbers, and the crosses are averaged number densities for each 5 Mpc. The top-right and bottom-left panels show relationships of relative flux versus exploding distance due to the production mass of $^{56}$Ni and viewing angle of a WD binary. The relation of the two types of flux variations is plotted in the bottom-right panel. The black filled circles, red open squares and green filled squares indicate SNe exploding at distances in the ranges of 0–20, 20–40, and 40–60 Mpc, respectively.

Additionally, we also estimated the level of sensitivity of a light curve for a given mixture ratio of SD and DD scenarios, as shown in Fig. 11. Our results show us that the ETCC satellite could determine the coexisting ratio of SD and DD scenarios with a 20%–30% uncertainty.

Similarly, we tried to extract the essential differences in averaged line spectra between SD and DD scenarios, as shown in Fig. 7. An essential difference between the SD and DD is an amount of peripheral mass of the exploded SN Ia. This means that in the DD scenario, low-energy emission lines are expected to be absorbed more strongly than those in the SD scenario (Summa et al. 2013). To estimate this effect, the averaged gamma-ray spectra are divided into four periods in 20–100 days after the explosion. Then, we plot the observable counting ratios of the line of 1.238 MeV and the energy ranges of 1.7–4.0 MeV for the SD and DD, as shown in Fig. 12. The results evidently indicate the differences in this ratio at each period.

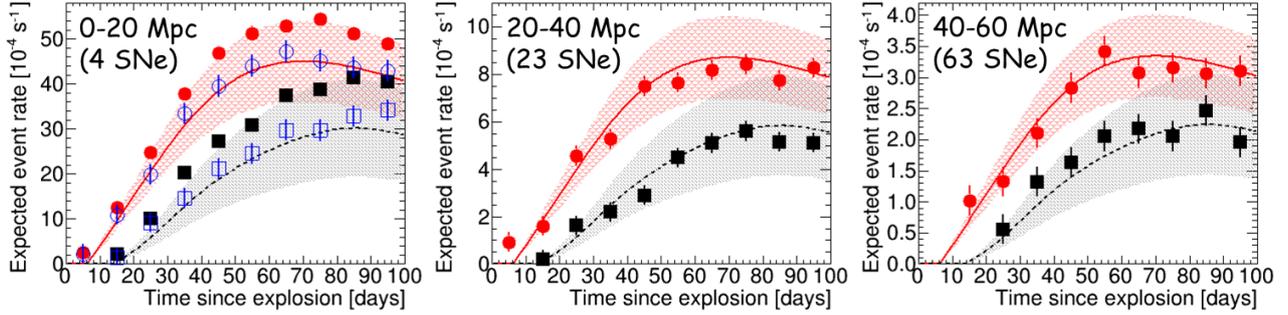

**Fig. 10,** Averaged light curves of SNe Ia exploding in the distance ranges of 0–20, 20–40 and 40–60 Mpc over five years. The filled circles and filled squares are averages of expected light curves by observing with the ETCC satellite for the progenitor scenarios of SD and DD, respectively. Fluctuations of intrinsic intensity and photon statistics are considered. The solid and dotted lines indicate assumed light curves without any fluctuations for SD and DD scenarios, respectively, and the hatched region show intrinsic fluctuation ranges. In the case for light curves within the range of 0–20 Mpc, the number of expected light curves is evidently higher than the number of assumed light curves. It can be explained by the individual property of a SN Ia with higher mass production of $^{56}$Ni. The open circles and open squares are also averaged light curves except for the SN with a high mass production of $^{56}$Ni in the 0–20 Mpc range. Such individual properties can be cancelled-out by averaging over the large number of SNe.

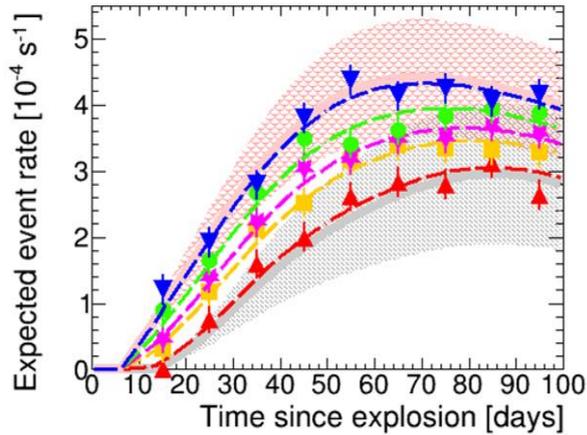

**Fig. 11,** Averaged light curves of SNe Ia in the distance range of 20–60 Mpc with a varying mixture ratio of DD and SD scenarios. The red triangles, orange squares, magenta stars, green circles, and blue inverted-triangles indicate the mixture ratio of the SD scenario with 0, 0.25, 0.5, 0.75, and 1, respectively. Each dashed line accompanying the plots is the best fitted light curve with a mixture of the SD and DD scenario.

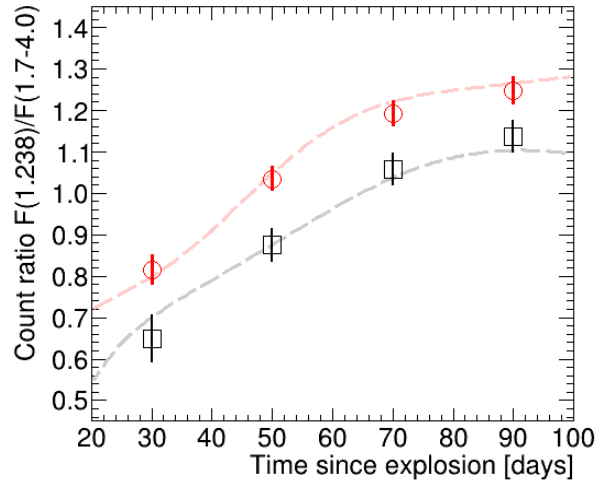

**Fig. 12**, Expected time evolution of hardness ratio of 1.238 MeV line to 1.7–4.0 MeV band, which are averaged over SNe Ia in the distance range of 20–60 Mpc. The red circles and black squares indicate scenarios of SD and DD, respectively.

Thus, statistical studies of MeV gamma-ray light curves of both the total bolometric emission and hardness ratio, which are realized by imaging spectroscopy, would provide a conclusive solution to the puzzle of the origin of SNe Ia.

Conclusions

Imaging spectroscopic observations are available for all electromagnetic wavelengths outside the Compton scattering region since the direction of a ray or photon can be mapped individually onto the imaging plane via antennae, lenses, reflectors or measuring the direction of each photon directly (GeV and TeV gamma-rays). Thus, we can certainly estimate the sensitivity from the detection area (effective area) and a well-defined PSF. However, conventional imaging methods such as CC for the Compton scattering region cannot provide individual mapping since CC measures only one of two angles defining a photon's direction, which makes it very difficult to estimate the true sensitivity. Figure 13 shows the schematic explanation of difference between the imaging spectroscopy of CC and ETCC. On the imaging plane of CC, the region of interest always includes some photons permeated from the outer area extending out of the FoV, and we eventually have to subtract these photons distributed in a wide area even if we know all the information regarding background photons. Therefore, the PSF of CC effectively extends up to the level of the Compton scattering angle (Tanimori et al. 2017). Furthermore, the background distribution of photon is usually unknown and needs the optimization of background modeling, such as the maximum-likelihood expectation maximization method for the background subtraction, which often causes larger systematic errors than statistical errors. In short, CC cannot observe intensity—a fundamental physical quantity in optics; hence, imaging spectroscopic observation is impossible for CC. However, ETCC can observe intensity rigidly and gives us a proper imaging spectroscopic observation in the Compton scattering region. The influence of imaging spectroscopy is evident in the results presented in this study.

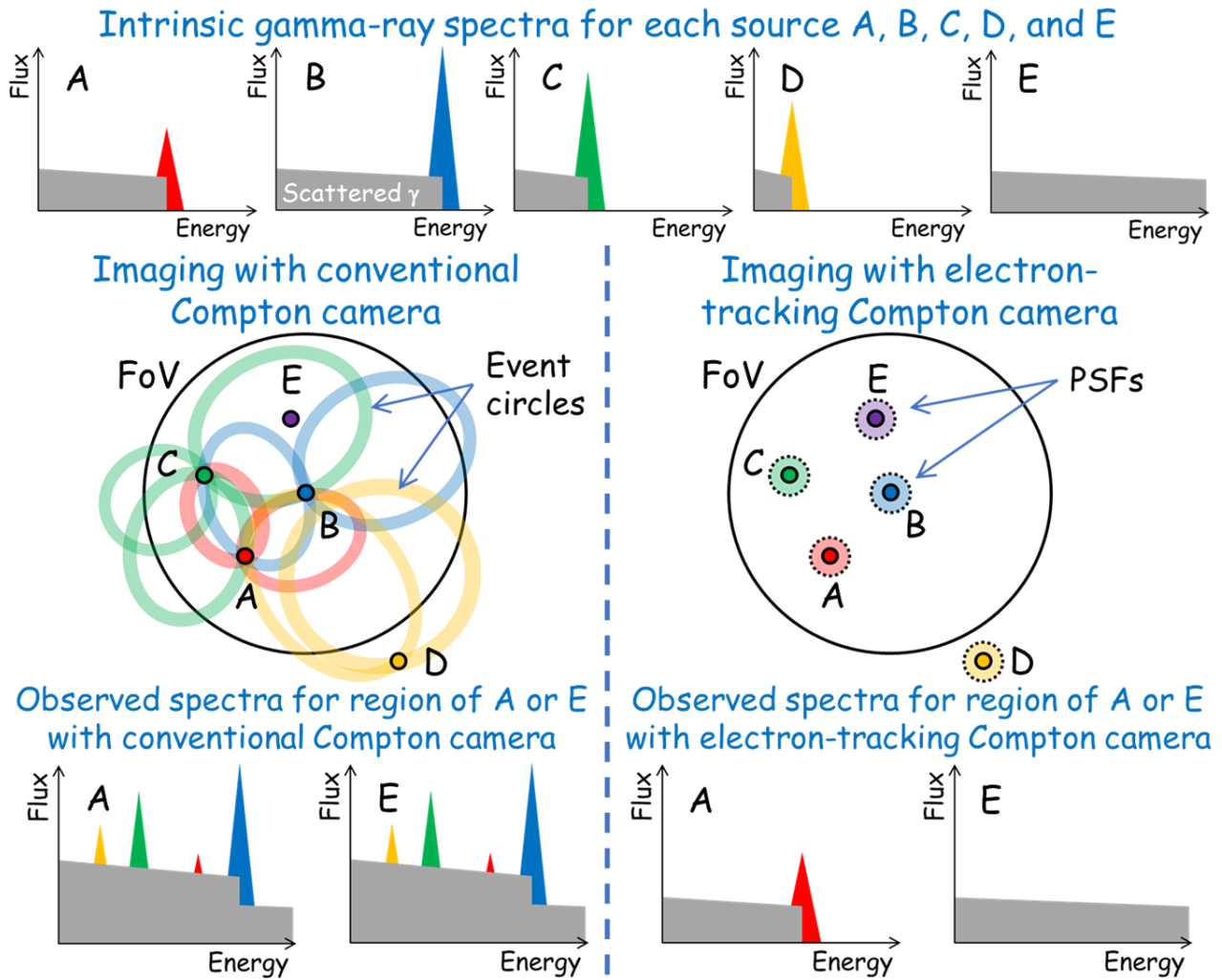

**Fig. 13**, Schematic explanation for imaging differences between CC and ETCC. In this picture, we assumed five point sources of A, B, C, D, and E. These have energy peaks, with the exception of source E. Only source D is located out of the FoV. In the case of imaging with conventional CC, reconstructed gamma-ray photons have annulus error regions. However, in the case of imaging with ETCC, reconstructed gamma-ray photons are concentrated at the PSFs region. Although imaging spectroscopy by CC for region of a source is heavily affected by the other sources, ETCC technology can suppress the spectroscopic effect of contamination from other gamma-ray sources.

ACT (ACT study team 2005; Boggs 2006), which had the highest performance and the highest cost among all MeV gamma-ray astronomy projects implemented in this century, proposed the spectroscopic detection of SN Ia up to 20 Mpc. GRIPS, a moderate project proposed in 2014, was expected to measure the spectrum of SNe up to ~6 Mpc (Summa et al. 2013). The performances of recent projects were lower than that of GRIPS owing to the use of a crystal (CsI) with lower energy resolution than that of $LaBr_3$, which was the crystal used in GRIPS. Although the effective areas of the ETCC satellite and GRIPS are similar (~200 $cm^2$ at 1 MeV), their expected number of spectroscopic observations of SN Ia within five years of operation are ~100 and 1–2, respectively. Notably, the advent of Einstein, which provided proper imaging, and ASCA, which provided proper imaging spectroscopy, advanced X-ray astronomy dramatically. ETCC technology gives MeV gamma-ray astronomy these two benefits at once. With ETCC technology, MeV gamma-ray astronomy will reach the sub-mCrab sensitivity region at above 0.1 MeV for the first time in high energy astronomy.

A PSF above 5 MeV also provides a new possibility to explore the multi-MeV region. In this energy region, the cross section of the pair creation gradually exceeds that of Compton scattering. However the directional resolution of the pair creation is intrinsically limited to above 0.1 degrees for energies lower than 100 MeV (1° at 10 GeV) by an unknown momentum transfer to the nuclei. Realistic resolution of future detectors will most likely be 10° to 1° between 10 and 100 MeV. Then at up to 50 MeV, a sensitivity improved by a factor of 10 will be made possible by the ETCC satellite, even by considering the degradation of the cross section of Compton scattering.

Only gas tracking technology can provide a fine tracking of a recoil electron necessary for sub-degree PSF, and the adoption of a well-defined PSF is a unique method to explore and advance MeV gamma-ray astronomy. In April 2018, SMILE-II+ was launched by the JAXA balloon team and observed 2/3 of the whole sky for 26 hours from Alice Springs, Australia. We will soon provide results demonstrating the abilities of imaging spectroscopic observation for MeV gamma-ray astronomy.

Acknowledgments

We deeply appreciate the fruitful discussions about SNe with Prof. Keiichi Maeda at Kyoto University and Dr. Alexander Summa at Würzburg University. This study was supported by the Japan Society for the Promotion of Science (JSPS) Grant-in-Aid for Challenging Exploratory Research (25610042), Grant-in-Aid for Young Scientists (B) (15K17608), and a Grant-in-Aid from the Global COE program, Next Generation Physics.